\documentclass[twocolumn,showpacs,preprintnumbers,amsmath,amssymb]{revtex4}
\usepackage{graphicx}
\usepackage{dcolumn}
\usepackage{bm}
\usepackage{color}

\newcommand {\be}{\begin{equation}}
\newcommand {\ee}{\end{equation}}
\newcommand {\bea}{\begin{eqnarray}}

\newcommand {\eea}{\end{eqnarray}}

\begin{document}

\title{High-density kaonic-proton matter ({\it KPM}) 
composed of $\Lambda^* \equiv K^-p$ multiplets\\
 and its astrophysical connections \\}

\author{Yoshinori Akaishi$^{1,2}$ and Toshimitsu Yamazaki$^{1,3}$}

\affiliation{$^{1}$ RIKEN, Nishina Center, Wako, Saitama 351-0198, Japan}
\affiliation{$^{2}$ High-Energy Accelerator Research Organization (IPNS/KEK), 1-1 Ooho, Tsukuba, Ibaraki 305-0801, Japan}
\affiliation{$^{3}$ Department of Physics, University of Tokyo, Hongo 7-3-1, Bunkyo-ku, Tokyo 113-0033, Japan}

\thanks{} 
\date{\today, KPM-PRL-DAE-v4}

\begin{abstract}
  We propose and examine a new high-density composite of  
    $\Lambda^* \equiv K^-p = (s \bar{u}) \otimes (uud)$, which may be called {\it Kaonic Proton Matter (KPM)}, or simply, {\it $\Lambda^*$-Matter}, where substantial shrinkage of baryonic bound systems originating from the strong attraction of the $(\bar{K}N)^{I=0}$ interaction takes place, providing a ground-state neutral baryonic system with a huge energy gap. 
      The mass of an ensemble of $(K^-p)_m$, where $m$, the number of the $K^-p$ pair, is larger than $m \approx 10$, is predicted to drop down below its corresponding neutron ensemble, $(n)_m$, since the attractive interaction is further increased by the Heitler-London type molecular covalency, as well as by chiral symmetry restoration of the QCD vacuum.   Since the seed clusters ($K^-p$, $K^-pp$ and $K^-K^-pp$) are short-lived, the formation of such a stabilized relic ensemble, $(K^-p)_m$, may be conceived during the Big-Bang Quark Gluon Plasma (QGP) period in the early universe before the hadronization and $quark$-$anti$-$quark$ annihilation proceed.  
           At the final stage of baryogenesis a substantial amount of primordial ($\bar{u},\bar{d}$)'s are transferred and captured into {\it KPM}, where the anti-quarks find places to survive forever. 
      The expected {\it KPM} state may be {\it cold, dense and neutral $\bar{q} q$-hybrid ({\it Quark Gluon Bound (QGB)}) states, $[s(\bar{u} \otimes u) ud]_m$,} to which the relic of the disappearing anti-quarks plays an essential role as hidden components.  Explosive production of {\it KPM} from supernova precursors is considered as a possible observational astronomical process. 
\end{abstract}


\maketitle

\noindent
 {\bf Introduction}
  
  In the present paper we propose and examine a new high-density neutral matter, {\it anti-Kaonic Proton Matter (KPM)}, composed of hitherto known units of
\begin{equation} 
\Lambda^* \equiv K^-p  = (s \bar{u}) \otimes (uud),
\end{equation} \label{e1}
\noindent
which may be called {\it KPM}, or simply, {\it $\Lambda^*$-Matter ($\Lambda^*$-M)}. Its free unit, $\Lambda^*$, first predicted by Dalitz and Tuan \cite{Dalitz:59}, has been identified to be a known resonance state of $\Lambda(1405)$ with a mass of $M = 1405$ MeV/$c^2$ \cite{PDG14}. Its spectacular nature was not fully realized before.

The present investigation arises from our recent theoretical finding of high-density anti-{\it Kaonic} ($\bar{K}$) few-body {\it Nuclear Clusters} ({\it KNC}) \cite{akaishi:02,yamazaki:02,dote:04,yamazaki:04,yamazaki:07a,akaishi:09,yamazaki:07b,maeda:13}, where nuclear systems with a density of $\rho \approx 3 \rho_0$ ($\rho_0$ being the normal nuclear density, 0.17 fm$^{-3}$) are spontaneously formed, driven by the strong $(\bar{K}N)^{I=0}$ attraction without the aid of gravity. 
 We start our discussion from empirical information concerning the most important building blocks: $K^-p~ (= \Lambda^*)$, $K^-pp$ and $K^-K^-pp$. \\
 
i) Recent observations \cite{yamazaki:10,ichikawa:15} of the predicted dense state of $\Lambda^*$-$p \approx K^-pp$ \cite{yamazaki:07b}, which is the simplest form of {\it KNC}, support the theoretical framework for dense kaonic nuclear bound states \cite{yamazaki:07a,yamazaki:07b,akaishi:09,maeda:13}. 

 ii) Furthermore, a recent analysis \cite{akaishi-CLAS} of high-precision measurements of photo-induced reaction $p(\gamma, K^+)\Sigma^0\pi^0$ at CLAS \cite{CLAS} has yielded a precise value for $M (\Sigma^0 \pi^0)$, which reconfirmed the traditional value of the $\Lambda(1405)$ resonance mass \cite{PDG14} ($1405.1 ^{+1.3} _{-1.0}$ MeV/$c^2$) that favors the strong $(\bar{K}N)^{I=0}$ attraction in contrast to the prevailing double-pole hypothesis, which claims a much weaker attraction with a mass of $M \approx 1420$ MeV/$c^2$ \cite{Jido:03,Mai-Meissner}. 

iii) In $K^-pp \sim \Lambda^* p$ and $K^-K^-pp \sim \Lambda^*  \Lambda^*$
a molecular analogy stands even for the systems of nuclear interactions \cite{yamazaki:07a}, and the Heitler-London type covalent bonding effect \cite{Heitler-London} plays an important role as wide-ranging multiple bonding forces \cite{akaishi:09}.

iv) The spontaneous nuclear shrinkage causes an enhancement of the $\bar{K}N$ interaction by Chiral Symmetry Restoration (CSR) that iterates further production of higher nuclear densities,  and thus of larger kaonic binding energies and decreased masses of the {\it KPM} ensemble.

v) Thus, the joint effect of the multiple bonding of $\Lambda^*$ and the CSR  may cause a large energy gap, where the ground state of the $\Lambda^*$ multiplet may become well below that of the corresponding neutron  ensemble:
\begin{equation}
M[(K^-p)_m]~{\rm per~baryon} < M[(n)_m]~{\rm per~baryon}.
\end{equation}
\label{e2}

\noindent
 {\bf Multiple bonding of $\Lambda^* = K^-p$}
 
The double kaonic cluster, $K^-K^-pp$, initially predicted by \cite{yamazaki:04}, shows a well developed deeply bound structure of two $\Lambda^*$'s, whereas they persist to keep the identification as $\Lambda^*(=K^-p)$. Here, we comment on the interaction of the two $\Lambda^*(=K^-p)$'s. The original migrating exchange force of Heitler and London \cite{Heitler-London} was considered between two fermionic electrons in H$^+$-H$^0$ and H$^0$-H$^0$ molecules. In the present case, on the contrary, the migrating particles are {\it bosonic} $K^-$ mesons, the wave function of which is 
\begin{equation}
\Phi(\vec r_1,\vec r_2)=N(D)[\phi_a(\vec r_1)\phi_b(\vec r_2) +  \phi_b(\vec r_1)\phi_a(\vec r_2)],
\label{e3}
\end{equation}
where the two protons sit on sites $a$ and $b$ which are separated by a distance of $D$. Then, the exchange interaction is obtained as 
\begin{eqnarray}
&& \Delta U(D) \equiv U(D) - U(\infty) \approx 4\, \vert N(D)  \vert^2 \times\\ 
&& 
[\langle \phi_a \vert V_{K^-p} \vert \phi_b \rangle \langle \phi_b \vert \phi_a \rangle + \langle \phi_b \vert V_{K^-p} \vert \phi_a \rangle \langle \phi_a \vert \phi_b \rangle],
\label{e4}
\end{eqnarray}
where $U(D) = \langle \Phi \vert \sum^{i=1,2}_{j=a,b} V_{K^-_ip_j} \vert \Phi \rangle$ with an effective $\bar KN$ interaction, $V_{K^-p}$, as given by a $g$-matrix in \cite{akaishi:02}. 

\begin{figure}[htb]
\vspace{0cm}
\includegraphics[width=0.4\textwidth]{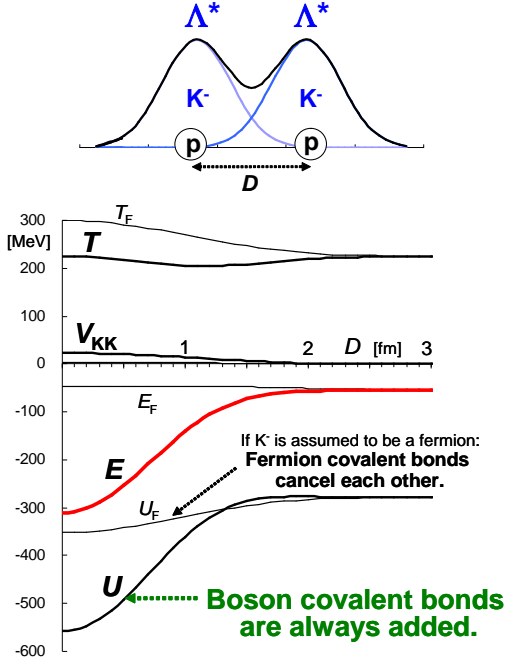}  
\vspace{-0.5cm}
\caption{\label{fig:L*L*model}  
A di-$\Lambda^*$ system. The exchange interaction of the two bosonic $K^-$'s cause large attraction compared with the case of assumed fermions. $U$, $T$, and $E$ stand for the potential, kinetic and total energies of two $K^-$'s.}
\end{figure} 

This $\Delta U(D)$, shown in Fig.~\ref{fig:L*L*model}, is a bonding potential due to doubly migrating $K^-$'s, and is about twice as strong as the one from single $K^-$ migration in $K^-pp$, discussed in \cite{yamazaki:07a}. On the other hand, if we artificially assume the migrating particles to be spinless fermions, the two terms of Eq.~(\ref{e3}) should be subtracted, and would yield a much weaker bonding. It is noted that the bonding from multi-$K^-$ migrations is always additively constructed due to the bosonic nature of $K^-$. In this way, the $K^-$'s bring about a much stronger binding effect. We have obtained the effective potential between the two $\Lambda^* (=K^-p)$'s  by folding the bonding potential, $\Delta U$, the $K^-K^-$ repulsive potential, $V_{KK}$, and a realistic $NN$ potential having a repulsive core, with the internal $K^-p$ distribution of $\Lambda^*$.
 
\begin{figure}[htb]
\vspace{0.5cm}
\includegraphics[width=0.47\textwidth]{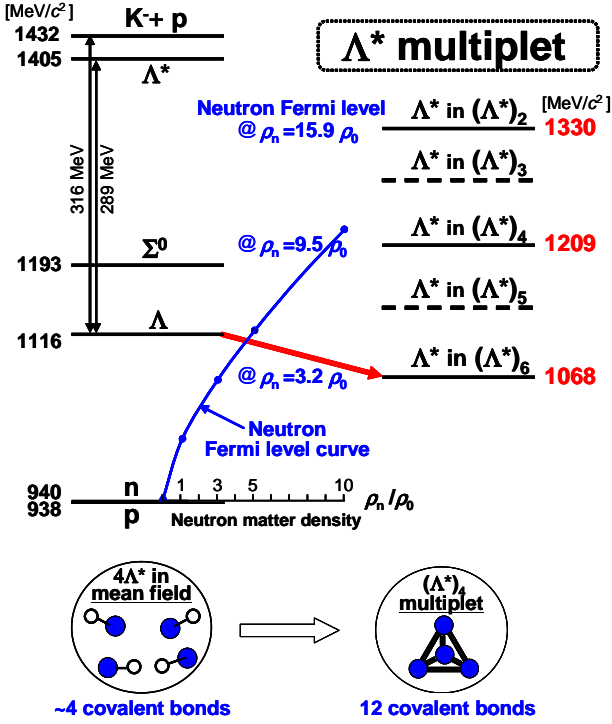}
\vspace{0cm}
\caption{\label{fig:$L^*$ levels} 
Predicted energy levels of $\Lambda^* = K^-p$ in $\Lambda^*$ multiplets calculated by a variational method \cite{ATMS}. The corresponding nuclear densities and neutron Fermi levels are also shown, indicating that the $\Lambda^*$ in the $(\Lambda^*)_6$ cannot decay to a neutron in neutron matter at 3.2 times the normal density $\rho_0$.} 
\end{figure}

We applied the effective $\Lambda^*$-$\Lambda^*$ interaction thus obtained to calculate the binding energies of multiple $(\Lambda^*)_m$ systems that approximate multiple $(K^-p)_m$ states. Here, we take into account the possible combinations of $\Lambda^*$-$\Lambda^*$ bonding, as the number of bonding increases with the multiplicity being $2, 6, 12, 20, 30,..,$  for $m  = 2, 3, 4, 5, 6,..,$ respectively. The results obtained by using a variational method (ATMS \cite{ATMS} employed in \cite{yamazaki:07a}) are shown in Fig.~\ref{fig:$L^*$ levels}, which indicate that the energy level per each $\Lambda^*$ of a multiple $(\Lambda^*)_m$ state drops down, and finally exceeds the threshold level of free $\Lambda$ emission, when the $\Lambda^*$ multiplicity becomes larger than some critical number. The number is estimated to be 10, if the effect of CSR (discussed in the next section) that will enhance the assumed basic $\bar KN$ interaction is taken into account. Such a multiplet as $(\Lambda^*)_{m>10}$ could be stable against any strong-interaction decay. 

\begin{figure}[tbh]
\begin{center}
\includegraphics[width=0.45\textwidth]{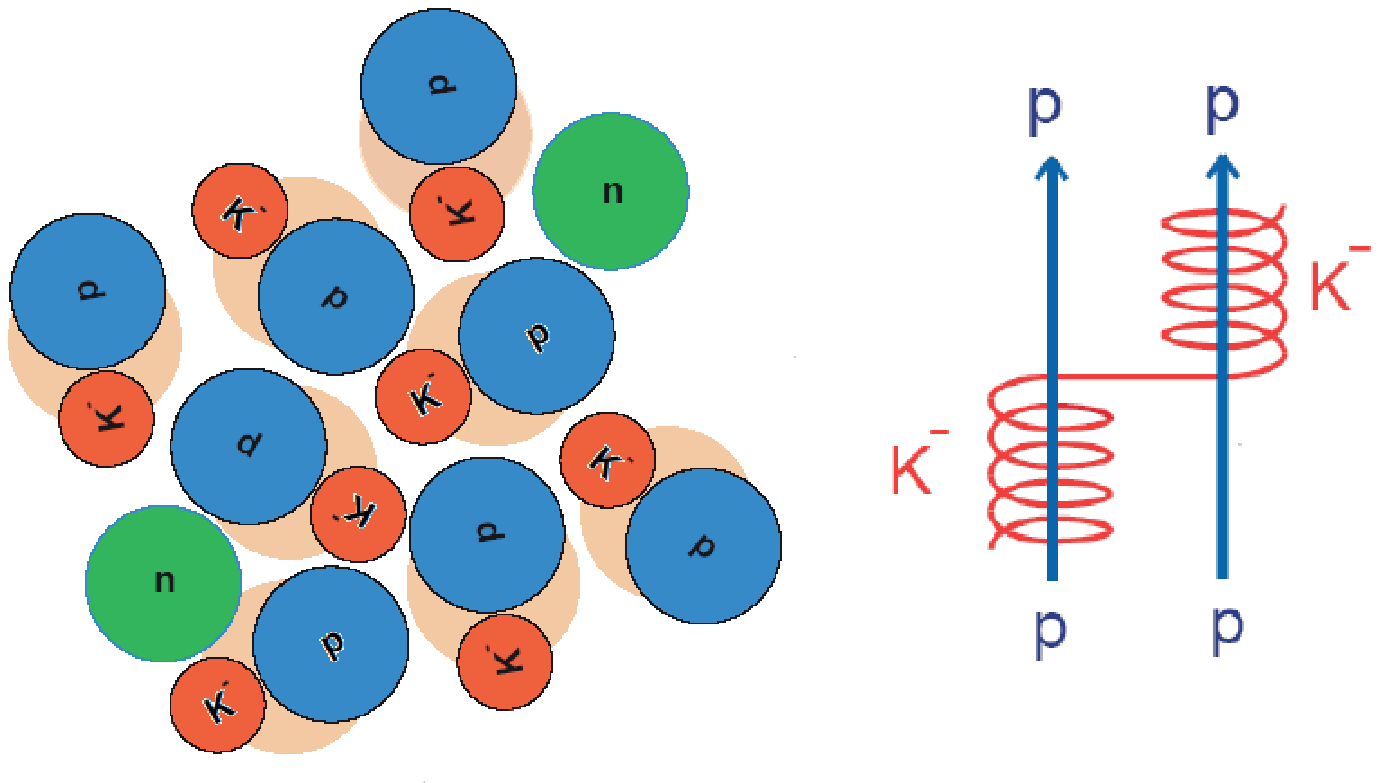}
\includegraphics[width=0.35\textwidth]{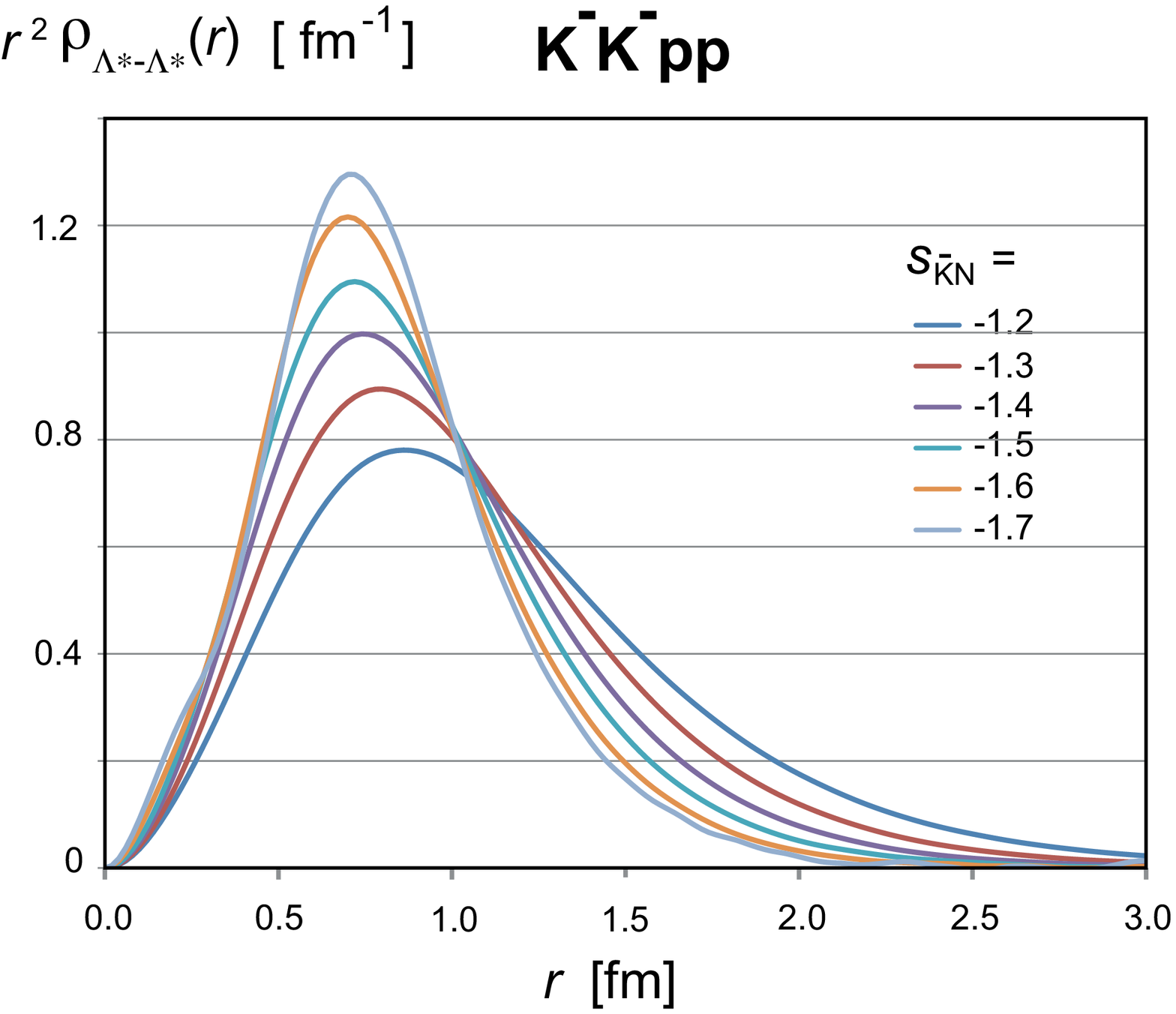}
\caption{(Upper) An ensemble of $\Lambda^*$'s together with Heitler-London type covalent bonding of bosonic $K^-$.  (Lower) Calculated distributions of the $\Lambda^*$-$\Lambda^*$ distance in $K^-K^-pp$ at various $\bar{K}N$ strength parameters. From \cite{maeda:13}. }
\label{fig:f3}
\end{center}
\end{figure}

Figure~\ref{fig:f3} shows a stable ensemble of $\Lambda^*$'s together with  Heitler-London type covalent bonding of bosonic $K^-$, that produces super-strong nuclear interaction \cite{yamazaki:07a}.  A mean-field model for multi $\bar{K}$fs in nuclei is employed in \cite{gazda:08}, but lacks just this multi-bonding mechanism of the super-strong nuclear attraction, which gives a drastic non-linear decrease of $M[(K^-p)_m]$ as $m$ increases. It should be mentioned that the $K^-$ in a nucleus cannot keep to hold its independent-particle motion in mean field by yielding a marked ($\Lambda^*=K^-p$) cluster correlation. In fact, the $K^-$ in a nucleus does not satisfy the ghealingh condition for independent-particle motion discussed by Gomes {\it et al.} \cite{gomes:58}. 

In order to see the effect of CSR on the size of the basic $\Lambda^* \Lambda^* $ system, distributions of the $\Lambda^*$- $\Lambda^*$ distance obtained from Faddeev-Yakubovsky calculation for $K^-K^-pp$ \cite{maeda:13} is also shown in Fig.~\ref{fig:f3}. 

\vspace{0.5cm}
\noindent
{\bf Stability and stiffness of $KPM$}

Here, we consider the basic stability of $KPM$. 
The longevity of $KPM$ depends on its stiffness against the addition of external foreign substances and the subtraction of internal components. 
The total mass of the $\Lambda^*$ multiplet in the preceding section, $M[(\Lambda^*)_m]$ per baryon, is well approximated as
\begin{equation}
M[(\Lambda^*)_m]c^2 \approx ~ m~1405_{\rm [MeV]} 
+ \frac{m(m-1)}{2} \langle \Delta U \rangle_{av.}
\label{e6}
\end{equation}
with $\langle \Delta U \rangle_{av.}=-135$ MeV for $m =4 \sim 8$. 
Then, the $\Lambda^*$ separation energy, $S_m(\Lambda^*)$, for the $(\Lambda^*)_m \to {\rm free}~\Lambda^* + (\Lambda^*)_{m-1}$ process is given by 
\begin{equation}
S_m(\Lambda^*) = - (m-1)~\langle \Delta U \rangle_{av.}.
\label{e7}
\end{equation}
It is noted that $S_m$ is 2-times larger than  $BE_m$ (binding energy per $\Lambda^*$), that is the mass difference between a free $\Lambda^*$ and a bound $\Lambda^*$ in $(\Lambda^*)_m$, due to a rearrangement of the $(\Lambda^*)_{m-1}$ cluster. The $S_6(\Lambda^*)$ is estimated to be 675 MeV, which is almost 2-orders of magnitude larger than the nucleon separation energy of about 8 MeV from usual nuclear systems. $S_m(\Lambda^*)$ becomes larger with $m > 6$.

As for the weak decay of $(\Lambda^*)_m$, the $Q$-value of the $(\Lambda^*)_m \to n + (\Lambda^*)_{m-1}$ non-leptonic process is given by 
\begin{equation}
Q_m(\Lambda^* \to n)=(1405-940)_{\rm [MeV]}+(m-1)~\langle \Delta U \rangle_{av.}.
\label{e8}
\end{equation}
In the case of $m=6$, the weak decays of single $\{ \Lambda^* \to n \}$ and also $(2 \sim 4) \times \{ \Lambda^* \to n \}$ are prohibited kinematically due to negative $Q$-values, though the mass of $\Lambda^*$ in $(\Lambda^*)_6$ is still heavier than the neutron mass, as shown in Fig.~\ref{fig:$L^*$ levels}. Only $(5 \sim 6) \times \{ \Lambda^* \to n \}$ take place through simultaneous weak decays, which are profoundly suppressed by the decay multiplicity.

Similarly, the kaon weak-decay $\{ K^- \to e^-+\bar \nu \}$ process of $(\Lambda^*)_m \to e^-+\bar \nu + p + (\Lambda^*)_{m-1}$ is strongly suppressed at $m=6$ and is prohibited at $m \geq 8$.  \\

\noindent\\
 {\bf Chiral symmetry restoration for $\bar{K}N$}
   
The recent experimental data on $K^-pp$ from DISTO \cite{yamazaki:10} and J-PARC E27 \cite{ichikawa:15} gave a binding energy of about 100 MeV, which is a factor of 2 larger than the original prediction \cite{yamazaki:02} based on the empirical $\Lambda(1405)$ mass. A recent Faddeev-Yakubovsky calculation \cite{maeda:13} shows that the observed binding energy corresponds to an effective $\bar{K} N$ interaction which is about 17\% more attractive than that assumed in the original prediction. Here, we consider the origin of this enhanced interaction in terms of the chiral-symmetry restoration (CSR) effect \cite{nambu:61,hatsuda-kunihiro:85,Vogel-Weise:91,Weinberg:66,GOR:66,BKR:87,yamazaki:12}.

\begin{figure}[htb]
\vspace{0.cm}
\includegraphics[width=0.4\textwidth]{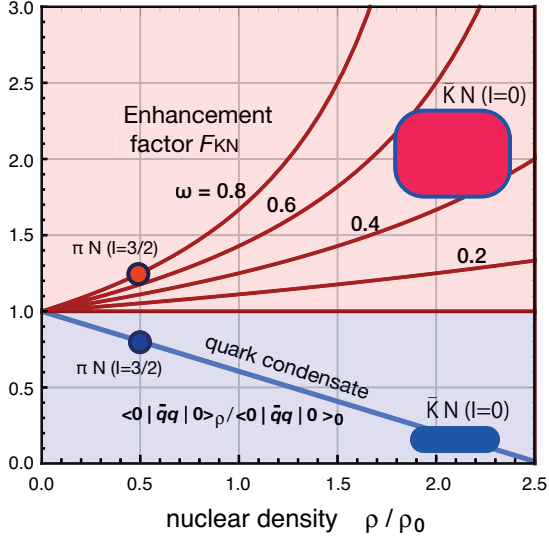}
\caption{\label{fig:CSR} 
 Chiral symmetry restoration effect of  $\pi$ and $\bar{K}$ (see details \cite{maeda:13}). (Lower half) The QCD condensate $|\langle\bar{q} \, q\rangle|$ decreases with the medium nuclear density $\rho$ in units of $\rho_0$.   
(Upper half) The enhancement factors by CSR, $F_{\bar{K}N}$, for various $\omega$ values of the vacuum clearing factor \cite{BKR:87}.}
\end{figure} 

In general, when CSR takes place in dense nuclear medium, the quark condensate decreases toward zero, and the  
 $(\bar{K} N)^{I=0}$ interaction is expected to increase in magnitude. A naive qualitative estimate was made in \cite{maeda:13} by employing a model of Brown, Kubodera and Rho (BKR) \cite{BKR:87}. Figure~\ref{fig:CSR} shows the estimated quark condensate (straight line) and the enhancement factors $F_{\bar{K}N^{I=0}}$ as functions of the nuclear density $\rho(r)$, where $\omega$ is a "QCD-vacuum clearing factor".
 In the case of $\bar{K} N^{I=0}$, a drastic situation takes place  \cite{maeda:13}; $F_{\bar{K} N}$ increases and amplifies the binding energy and shrinks the nucleus furthermore, leaving less and less room for the QCD vacuum with further increasing the $\omega$ and $F_{\bar{K} N}$ factor non-linearly.  
 An enhancement factor of 1.5 corresponding to the density $\rho / \rho_0 \approx 2$ produces an enormous multiplication of the binding energy of $K^-K^-pp$ \cite{maeda:13}. Although the above estimate is very rough, the CSR effect in combination with the Heitler-London type enhancement is expected to bring the {\it KPM} mass of moderate multiplicity ($m \sim 10$) well below the nucleon mass,.

\begin{figure}[htb]
\vspace{-0.5cm}
\includegraphics[width=0.5\textwidth]{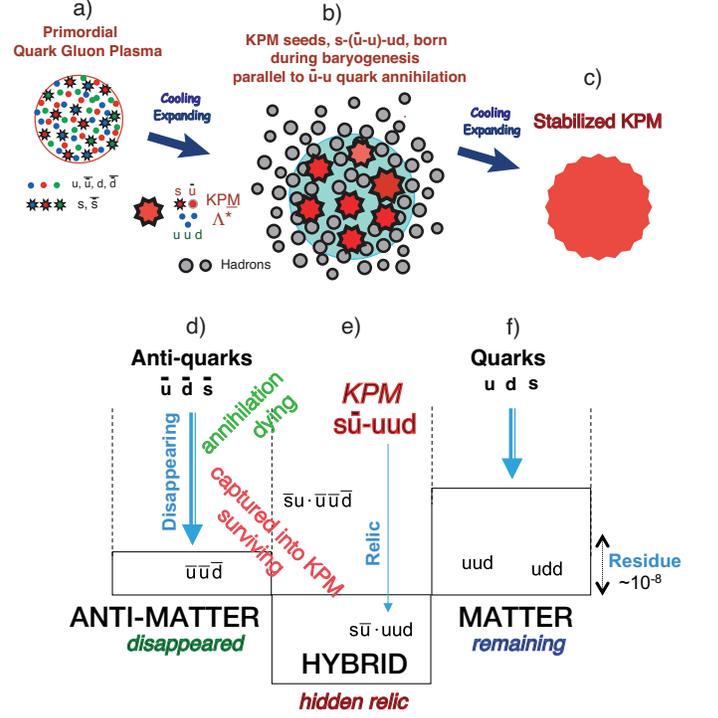}
\caption{\label{fig:QGP-KPM} 
(Upper) Formation of {\it KPM} from the primordial Big Bang (a) and (b), where $u, \bar{u}, d, \bar{d}, s, \bar{s}$ quarks are produced in QGP at high temperatures and densities. 
With  decreasing temperature it proceeds to the pre-hadronization stage, where $K^-p$, $K^-pp$ and $K^-K^-pp$ with large binding energies are formed, as indicated by the star-like red symbols. 
 Then, stable $(\Lambda^*)_{10}$ composites are formed, which  eventually grow larger and larger, but will become cold matter with eventual formation of $Quark$-$Gluon$-$Bound$~$(QGB)$ states. (Lower) Three quark sectors in the early universe during the disappearance of anti-quark matter: (d) the disappearing {\it ANTI-QUARK} sector, (e) quark-anti-quark {\it HYBRID} sector, where {\it relic and stable} precursors of $K^-p = s \bar{u} - uud$ are born, and (f) remaining ordinary {\it QUARK} sector.}
\end{figure}

\noindent\\
 {\bf How can {\it KPM} be formed?}
 
 As the {\it KPM seed} clusters, $K^-p$, $K^-pp$ and $K^-K^-pp$, are short-lived with $\Gamma \sim 100$ MeV, they cannot survive during the cascading collisions toward heavier clusters. Exceptional cases might take place during the initial phase of the early universe, where quarks ($u, d, s, \bar{u}, \bar{d}, \bar{s}$) and gluons are produced in quark gluon plasma (QGP) at extreme high temperatures and densities, but probably before the hadronization stage, as illustrated in Fig.~\ref{fig:QGP-KPM}. 

 Since the {\it KPM} seeds, particularly, $K^-p \equiv \Lambda^*$ and $K^-pp \equiv \Lambda^* p$, are distinctly deeply bound with binding energies of around 50 $\sim$ 100 MeV, whereas other quarks and hadrons are relatively shallowly bound, we expect that during the course of decreasing temperatures ($kT \approx 100$ MeV, and in expansion), the seeds are likely to become deep quasi-stable self-trapping centers, and recombined with other seeds that have just been born nearby.  The star-like red objects illustrated in Fig.~\ref{fig:QGP-KPM} (b) represent such just-born fresh composites of $\Lambda^*$ multiplets with $m \sim 10$. They undergo further combinations to become a large-scale more stable $KPM$. This process is in competition with 
   the branching ratio of $\Lambda^*$ formation $\{s \bar{u} + uub \rightarrow s (\bar{u} u) ud\}$ to the normal $\bar{q} q$ annihilation background, $\bar{q} + q \leftrightarrow g's$ in the early universe. Certainly, such competition occurs in the QCD level, and we need more knowledge on its answer.  

 It is to be noted that the basic unit of $KPM$, $\{(s\bar{u})  (uud)\}$, involves a $\bar{u}$-$u$ pair, which is essential in producing this deeply bound system. {\it This system possesses one $\bar{u}$ quark per unit that has been transferred from the primordial QGP phase.} Figure~\ref{fig:QGP-KPM} (d), (e), and (f) shows symbolically (d) a disappearing {\it ANTI-MATTER} sector that involves originally unbound $\bar{q}$ before $\bar{q} q$ annihilation, and (f) a dominating {\it MATTER} sector of relative baryon density around $2 \times 10^{-8}$, resulting after $\bar{q} q$ annihilation and baryogenesis. During the anti-quark disappearing stage {\it relic and  stable} composites of $\{(s\bar{u})  (uud)\}$ are formed, and constitute a (e) {\it HYBRID} sector.  In other words,   a substantial fraction of anti-particles may remain being  hidden relic in the {\it KPM} phase as an unknown astronomical object. \\

\noindent
{\bf Formation of {\it Quark-Gluon Bound (QGB)} states}

Annihilating, but still surviving, anti-quarks contribute to forming seeds for {\it KPM}: 
$ \Lambda^* \equiv [s (\bar{u}u) ud].$
This particle-anti-particle hybrid state has very strong attractive interactions with surrounding similar species; thus, multiple $\Lambda^*$ states are composited and their mutual fusions take place {\it in a short time and on a large scale}, as if it occurred in a sudden phase transition. 

 The above $\Lambda^*$'s that are defined as $K^-p$ in the language of hadrons may be born directly from constituents of QGP from the beginning, but eventually become cooled so as to be changed  into the new phase: {\it Quark Gluon Bound} ({\it QGB}) states. While being cooled furthermore, its {\it QGB} phase may remain unchanged. Whether $KPM$ could form a macroscopic object or not, the possibility of $KPM$ fragments as low-temperature QGB states should be an extremely interesting problem, as  no such quark-gluon bound states at low temperatures have been experienced so far either empirically or theoretically.
\\

\noindent
{\bf Production of KPM from supernova explosions}

Finally, we consider possible population of  {\it KPM} in connection with neutron stars $[n]_m$, which is somewhat similar to kaon condensation as discussed by Kaplan and Nelson \cite{Kaplan-Nelson:86} and Brown {\it et. al.} \cite{Brown94}.  The neutron stars (NS) once produced may proceed to {\it KPM} in gentle multiple decay processes that occur slowly:
\begin{equation}
[n]_{\rm NS} \rightarrow [K^-p]_{KPM} + (\nu + \bar{\nu})'s.
\end{equation}   
On the other hand, precursors of supernova explosion may undergo explosive processes toward not only to neutron-star (NS) formation but also to {\it KPM} formation:
\begin{eqnarray}
&&[e^-, p, n]_{\rm supernova} \rightarrow \nu's + [n]_{\rm NS} + n's,\\
&&[e^-, p, n]_{\rm supernova} \rightarrow \nu's + [K^-p]_{KPM} + n's.
\end{eqnarray}
This latter process has never been considered nor observed. It may be an interesting process, as we may anticipate some astronomical observational signals.  

\noindent
{\bf Concluding remarks}\\
\hspace{5mm}Very recently, new experiments have been carried out to search for hadron production in extremely high-energy Pb + Pb collisions at LHC-ALICE \cite{ALICE,pbm}, where the most important precursor $K^-K^-pp$ toward {\it KPM} (see Fig.~\ref{fig:$L^*$ levels}) can be investigated. 
Such a precursor can also be produced in the reactions ($p + p \rightarrow \Lambda^* + \Lambda^* + K^+ + K^+$) at lab energies of around 7 GeV \cite{yamazaki:11,maeda:16}. One can also study the $(\Lambda^*)_m$ multiplets with moderate multiplicity, $m$, in heavy-ion reactions, which are expected to exist as metastable fragments with various lifetimes. They might include important $QGB$ fragments.
\\

The authors are grateful to Prof. Makoto Kobayashi for the illuminating discussion. This work is supported by the Grant-in-Aid for scientific work of Monbukagakusho.

\end{document}